\documentstyle[12pt,fleqn]{article}
\begin{document}
\begin{center}
Final version in \ Physics Letters A 203 (1995) 150--151\\[15mm]
{\Large Reply to the comment of Y.~Aharonov and L.~Vaidman on\\[2mm]
``Time asymmetry in quantum mechanics:\\[2mm]
a retrodiction paradox''}
\\[1cm] Asher Peres\\[7mm]
{\sl Department of Physics, Technion -- Israel Institute of
Technology, 32 000 Haifa, Israel}\\[1cm]
\end{center}

\noindent{\bf Abstract}\bigskip

In the standard physical interpretation of quantum theory, prediction
and retrodiction are not symmetric.  The opposite assertion
by some authors results from their use of non-standard interpretations
of the theory.\\[10mm]

In their criticism of my Letter [1], Aharonov and Vaidman [2] consider
the following situation: ``Assume that the $x$ component of the spin of
a spin-$1\over2$ particle was measured at time $t$, and was found to be
$\sigma_x=1$. The predictions and retrodictions for the results of
$\sigma_x$ measurements performed after or before the time $t$ are
identical. In both cases we are certain that $\sigma_x=1$.'' This
claim cannot be derived from standard quantum mechanics.

The correct statement is the following: we can, if we wish, perform
another measurement of $\sigma_x$ at a later time $t_2>t$, and then we
shall indeed find $\sigma_x=1$ (provided that $H=0$). On the other hand
we {\em cannot\/} decide at time $t$ to perform a measurement of
$\sigma_x$ at an {\em earlier\/} time $t_1<t$. At most, we can speculate
what would have been the result of such a measurement, if it had been
performed. This is counterfactual reasoning, and great care must be
exercised. All depends then of what we actually {\em know\/} of the
state of the system before $t$. If we know that it was an eigenstate of
$\sigma_x$, then obviously this was the eigenstate with $\sigma_x=1$. On
the other hand, if we know that it was an eigenstate of $\sigma_y$
(because the system happens to have been prepared in such a way some
time before $t_1$) the result of a measurement of $\sigma_x$ at time
$t_1$ is {\em random\/}. It is of course the same as the result of a
subsequent measurement of $\sigma_x$ at time $t$, but it would be
incorrect to assume that the latter must still be $\sigma_x=1$ if the
measurement at time $t_1$ is actually performed. Indeed, we are
considering now two different and mutually exclusive experimental
setups: measuring $\sigma_x$ at times $t_1$ and $t$, or only at time
$t$. There is no reason whatsoever to assume that the result obtained at
time $t$ is the same for these two different setups.

More generally, if an observer has a partial knowledge of the
preparation of a physical system, for example that there are a~priori
probabilities $p_m$ of having a density matrix $\rho_m$, and if an ideal
measurement of that system produces a definite pure state $\psi$, we can
use Bayesian statistics [3] to deduce posterior probabilities for
$\rho_m$, namely

$$ P(\rho_m|\psi)=p_m\,\langle\psi|\rho_m|\psi\rangle\Bigm/
 \sum_n p_n\,\langle\psi|\rho_n|\psi\rangle.            $$

\noindent This is the only conclusion that can be legitimately derived
from conventional quantum mechanics.

A formal statement of the above property is that an optimal
determination of the past of a system can be achieved by an
informationally complete set of physical quantities. Such a set is
always strongly noncommutative. On the other hand, an optimal
determination of the future of a physical system is obtained by a
Boolean complete set of quantities [4].

Aharonov and Vaidman [2] also write ``Peres encounters a retrodiction
paradox because he uses the standard approach \ldots'' I indeed stated
in [1] that I was following the conventional ``orthodox'' quantum
formalism, namely the one that is actually used by experimental
physicists for analyzing the results recorded by their instruments.
This standard interpretation is incompatible with the
``time-symmetrized'' version of quantum mechanics [5]. Obviously, a
different conclusion may be reached by those who prefer to use a
different ``interpretation'' of quantum theory, because that
``interpretation'' only is a convenient
euphemism for what amounts to an essentially different theory.

\begin{enumerate}
\item A. Peres, Physics Letters A 194 (1994) 21.
\item Y. Aharonov and L. Vaidman, Physics Letters A [preceding Letter].
\item C. W. Helstrom, Quantum detection and estimation theory (Academic
Press, New York, 1976).
\item P. Busch and P. J. Lahti, Found.\ Phys.\ 19 (1989) 633.
\item W. D. Sharp and N. Shanks, Philos.\ Sci. 60 (1993) 488.
\end{enumerate}

\end{document}